\documentclass[12pt,a4paper]{iopart}
\usepackage{times}
\usepackage{graphicx}
\usepackage{amssymb}
\begin{document}
\title[]%
{Vector exchanges in production of light  meson  pairs and elementary atoms.}
\author{S. R. Gevorkyan,  E. A.  Kuraev, M. K. Volkov}
\address{Joint Institute for Nuclear Research, 141980, Dubna, Russia }
\def \ba{\begin{eqnarray}}\def\ea{\end{eqnarray}}
\def\bc{\begin{center}}\def\ec{\end{center}}
\def\br{\left( \right)}
\def\nn{\nonumber\\}
\newcommand{\brs}[1]{\left[ #1 \right]}

\begin{abstract}
The production of  pseudoscalar and scalar mesons pairs  and  bound states (positronium  or pionium  atoms) in high energy $\gamma\gamma$ collisions at high energies  provided   by   photon or vector meson exchanges are considered. The vector exchanges lead to nondecreasing with energy cross section of binary    process    $\gamma+\gamma\to h_a+h_b$  with $h_a, h_b$   states in the fragmentation regions of initial particles.                                                                                                                                                                                                                                                                                                                                                         
The production of light mesons pairs  $\pi\pi, \eta\eta, \eta'\eta', \sigma\sigma $     as well as a pairs of   positronium  $Ps$ and pionium $A_\pi$ atoms in  peripheral kinematics are discussed. Unlike the photon exchange the  vector meson exchange needs a reggeization,  leading  to fall  with energy. Nevertheless due to peripheral  kinematics  out of very forward production angles the  vector meson exchanges dominated.\\
The proposed   approach allows to express  the  matrix elements of the considered processes  through  impact factors, which can be calculated in perturbation models like Chiral Perturbation Theory (ChPT)  or Nambu-Jona-Lasinio (NJL)   model  or determined from   $\gamma\gamma$ sub-processes or vector mesons radiative decay widths.\\
We obtain the cross sections for pionium atom production  in  collisions of high energy  pions  and electrons with protons. The possibility to measure these processes in experiment are discussed.
\end{abstract}

\maketitle

\vspace{1cm}

\section{Introduction}
The next  large project after LHC should be likely a linear $e^+e^-$  accelerator    at  energy $\sqrt{s}=0.5-1TeV$,  giving  exciting challenge  to study  $\gamma\gamma$  interactions at energies of hundreds GeV.  The technology of obtaining the beams of high energy photons is based on the backward Compton scattering of laser light on high energy electrons~\cite{telnov06}, idea known for  many years~\cite{arutjunian63,serbo10}.\\
Exclusive processes with hadronic final states test various model calculations and hadron production mechanisms.
So far the  meson pairs production  in two photon collisions are measured~\cite{lep1,lep2,kek1}  at   $\gamma\gamma$ center of mass energy  $W \leq 4$ GeV and  scattering angle $|cos{\theta}|<0.8$.\\
In this work we investigate  the production of  light mesons pairs  and  elementary atoms (positronium  Ps and pionium $A_\pi$ atoms) in high energy $\gamma\gamma$ collisions in peripheral kinematics:
\ba
\gamma(k_1)+\gamma(k_2)\to h_a(p_1)+h_b(p_2);~~~ h_a, h_b=\pi, \eta, \eta\prime, \sigma, Ps,  A_\pi
\ea
Due to  peripheral kinematics ( $ s=(k_1+k_2)^2,  t=q^2=(p_1-k_1)^2;~~~  s >> |q^2| $ )  the created objects $h_a, h_b $  have  energies approximately equal to the energies of  colliding photons  and move  along the  directions  of  initial particles motion  (center of mass of initial particles implied).\\
The dominant contribution to peripheral processes comes from large  orbital momenta in scattering amplitude expansion.
The background from low orbital momentum in peripheral kinematics is strongly suppressed  unlike the processes allowing  production at any angles. A  typical example  is the Born-term  amplitude ( $\pi$ exchange in the t channel) of  the process $\gamma\gamma\to \pi^+\pi^-$ whose differential cross section at small angles  has additional suppression due to wide phase volume of the final state.\\
The another remarkable property of the relevant cross sections-they become  independent from center of  mass  energy s of colliding particles starting from some threshold energy  $\sqrt{s}\sim 2-3 GeV$.  The nondecreasing feature of pairs yield is a result of vector nature of the interaction (photon  or vector meson exchanges in the t-channel (Fig.1)).\\
In peripheral kinematics one can use the perturbation models of hadrons like Chiral Perturbation Theory~\cite{weinberg79, gasser85} (ChPT) or Nambu-Jona-Lasinio~\cite{volkov05} (NJL) model to describe the sub-processes at the relevant vertexes. One can expresses  the matrix element of  reaction (1) through  so called impact factors, which are nothing else than the matrix elements  of  sub-processes (Fig.1a):
$\gamma(k_1)+\gamma^*(q)\to h_a $  and   $\gamma(k_2)+\gamma^*(q) \to h_b $ or (Fig. 1b):  $\gamma(k_1)+V(q)\to h_a $  and   $ \gamma(k_2)+V(q) \to h_b $ .\\

\begin{figure}[ht]
\begin{center}
\includegraphics[scale=0.5]{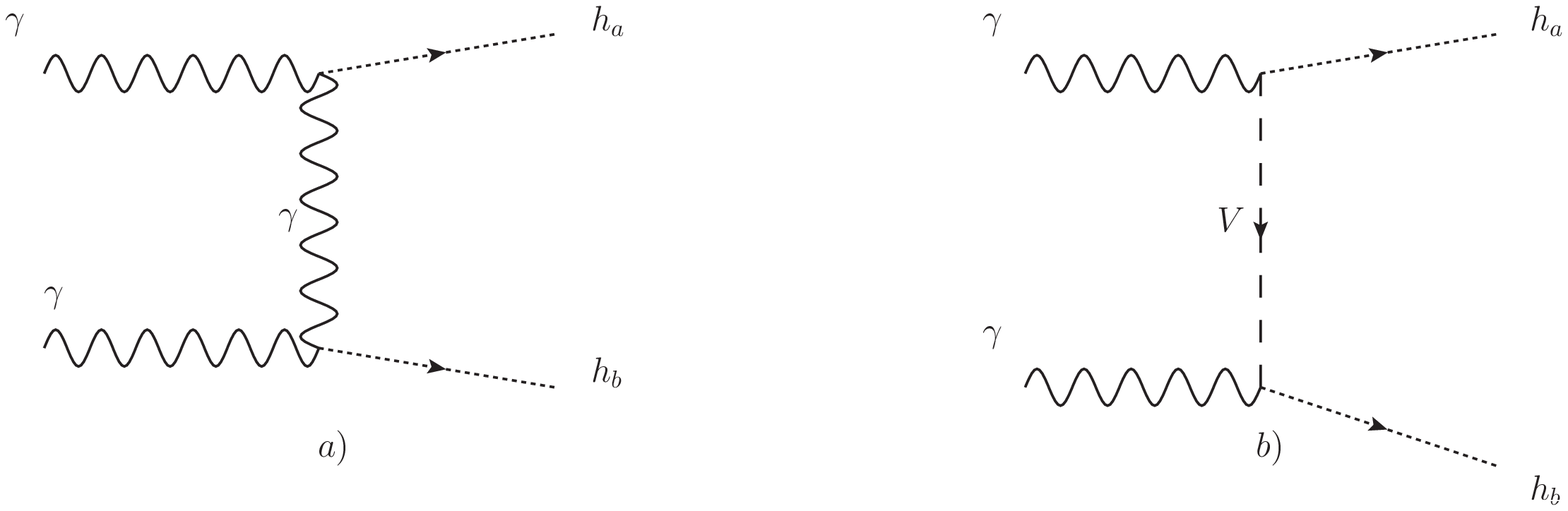}
\caption{ (a) The photon exchange in the process $ \gamma+\gamma\to h_a+h_b$ ;  (b) Exchange by vector meson.}
\end{center}
\end{figure}

Let us briefly discuss  the connection of matrix element  of reaction   (1)  with relevant impact factors  $M^a, M^b$  (the details can be found in~\cite{arbuzov10}). \\
According  to the  general rules the matrix element of  the process  (1) reads :
\ba
M=\frac{J^a_{\rho}J^b_{\sigma}}{q^2-m_V^2}g^{\rho\sigma},
\ea
 $J^{a,b}$ are currents associated with blocks $a,b$  of  relevant Feynman diagram.\\
Making use the infinite momentum frame parametrization of the transferred momentum:
\ba
q=\alpha k_1+\beta k_2+q_\bot;~~~q_\bot k_1=q_\bot k_2=0;~ q_\bot^2=-\vec q^2 <0; ~k_1^2=k_2^2=0.
\ea
and written metric tensor in the Gribov's  form:
\ba
g^{\rho\sigma}=g_\bot^{\rho\sigma}+\frac{2}{s}(k_1^\rho k_2^\sigma+k_2^\rho k_1^\sigma).
\ea
one  obtains the connection of matrix element of process (1) with impact factors  (with power accuracy):
\ba
M=\frac{2 s}{q^2-m_V^2} M^a M^b;~~~
M^a=\frac{J^a_\mu k_2^\mu}{s};~~~ M^b=\frac{J^b_\nu k_1^\nu}{s}.
\ea
The  impact factors  $M^a, M^b$   don't  decrease with energy  and can be described in terms  of perturbation  strong interaction models like Nambu-Jona-Lasinio model  or Chiral Perturbation Theory.
The cross section of the processes (1) is connected with  matrix element (5) in the standard way:
\ba
d \sigma^{a b \to h_a h_b}=\frac{1}{8s}\sum{|M|^2}d\Gamma
\ea
Expressing   the phase volume of  the two final particles $d\Gamma$  through the Sudakov parameters  (3)  one can rewritten  the two particles  phase space volume:
\ba
d\Gamma=(2\pi)^4\delta(k_1+k_2-p_1-p_2)\frac{d^3 p_1}{2(2\pi)^3E_1}\frac{d^3 p_2}{2(2\pi)^3E_2}
\ea
in the following form ~\cite{arbuzov10}:
\ba
d\Gamma=\frac{d^2q}{2(2\pi)^2s}
\ea
As a result the differential cross section of the processes (1) reads:
\ba
d \sigma^{a b \to h_a h_b}=\frac{d^2 q}{(4\pi)^2(\vec{q}^2+m_V^2)^2}\sum_{spins}|M^a|^2\sum_{spins}|M^b|^2
\ea
Thus the knowledge of relevant impact factors  allows one to calculate the cross sections of processes (1).
\section{Mesons  production. Photon exchange.}
We start with  the production of $\pi^0\pi^0$  pair in $\gamma\gamma$ collisions with photon exchange in the t-channel (Fig. 1a). The current algebra gives for the matrix element of neutral pion decay to two photons $ \pi^0(p)\to \gamma(k_1,e_1)+\gamma(k_2,e_2)$:
\ba
M(\pi^0\to\gamma\gamma)=\frac{\alpha}{\pi f_\pi}(k_1e_1k_2e_2),
\ea
where $(abcd)=a^\alpha b^\beta c^\gamma d^\delta \epsilon_{\alpha\beta\gamma\delta}$ and $k_i,e_i(k_i)$ are the
momenta and polarization vectors of real photons,  $\alpha=\frac{e^2}{4\pi}=1/137$ is the fine structure constant  and
$f_\pi=92.2 MeV$  is the pion decay constant measured in the $\pi^+\to\mu^+\nu_\mu$ decay rate.\\
The pion radiative decay width is given by the textbook formula:
\ba
\Gamma(\pi^0\to 2\gamma)=(\frac{m_\pi}{4\pi})^3(\frac{\alpha}{f_\pi})^2=7.76 eV
\ea
The decay amplitude (10) can be used as impact factor in  $\pi^0\pi^0$ production.\\
More elaborated  impact factors  considering the photon virtuality can be obtained if one calculates the triangle fermion loop with the light  u and d  quarks as a fermions~\cite{volkov05}.  Quarks charges and number of colors  result  in a factor $3((2/3)^2-(1/3)^2)=1$.  After standard procedure of denominators joining, calculating the relevant trace in the  fermions spin indices and  integration over the loop momenta  we obtain:
\ba
M(\pi^0\to\gamma\gamma^*)=\frac{\alpha}{2\pi f_\pi}|\left[\vec e,\vec q\right]|  F_\pi(z),~~~
  z=\frac{\vec q^2}{m_q^2} \nn
 F_\pi(z)=N_\pi\int\limits_0^1 d x\int\limits_0^1\frac{y d y}{1-\rho_\pi^2 y^2x(1-x)+z y(1-y)x},~~~
 \rho_\pi=\frac{m_\pi}{m_q}.
\ea
Here $m_q$  is the  constituent quark mass which we put $m_q=m_u=m_d\approx 280 MeV$, whereas  $N_\pi$ is the normalization constant  $F_\pi(0)=1$.\\
The similar expression for the sub-process of the scalar meson decay $\sigma\to\gamma\gamma^\ast $ reads:
\ba
M(\sigma\to \gamma\gamma^*)=\frac{5\alpha}{6\pi f_\sigma}|\left(\vec e,\vec q\right)| F_\sigma(z); ~~~ F_\sigma(0)=1,~~~ f_\sigma\approx f_\pi \nn
F_\sigma(z)=N_\sigma\int\limits_0^1 d x\int\limits_0^1\frac{y (1-4y^2x(1-x)) d y}{1-\rho_\sigma^2 y^2x(1-x)+z y(1-y)x},~~~
\rho_\sigma=\frac{m_\sigma}{m_q}.
\ea
The combination of quark charges  and color factor give a coefficient $3((2/3)^2+(1/3)^2)=5/3$. The nontrivial  difference in numerators of (12) and (13)  is a result of scalar nature of $\sigma$ meson.  As to the decay  $\eta(\eta')\to \gamma\gamma^*$    it is enough to do the relevant  replacements of masses in equation (12).\\
The amplitudes $M(\pi^0\to \gamma\gamma^*),~~~  M(\sigma\to \gamma\gamma^*)$  are nothing else than impact factors, one needs to calculate the cross sections of  neutral  mesons pairs production . Now we are in position to estimate the influence of the photon virtuality on the cross section of the reaction $\gamma\gamma\to\pi^0\pi^0$ from (1).  Making use the relation:
\ba
\int\limits_0^{2\pi}\frac{d\phi}{2\pi}[\vec{q}\vec{e}_1]^2[\vec{q}\vec{e}_2]^2=\frac{(\vec q^2)^2}{8}(1+2\cos^2\phi_0),
\ea
with  $\phi_0$  the  azimuthal angle between the initial photons polarization vectors. Substituting expression  (12)  for  $\pi^0$  impact factor in (9) we get:
\ba
\frac{d\sigma}{dz}=\frac{m_q^2}{8\pi}(\frac{\alpha }{4\pi f_\pi})^4 \frac{(z  F_\pi (z))^4}{ (1+z^2)^2}(1+2cos^2{\phi_0}).
\ea
In the case of pions production  one can safely neglect the small term $y^2x(1-x)m_\pi^2/m_q^2<0.05$ in the denominator of (12)  with the result:
\ba
F_\pi(z)=2\int\limits_0^1 d x\int\limits_0^1\frac{y d y}{1+z y(1-y)x}=\frac{4}{z}\ln^2\left(\sqrt{1+\frac{z}{4}}+\sqrt{\frac{z}{4}}\right).
\ea

The total cross section of the two neutral pions production:
\ba
\sigma^{\gamma\gamma\to \pi_0\pi_0}=\sigma_0(1+2\cos^2\phi_0)I, ~~~
\sigma_0=\frac{\alpha^4m_q^2}{2^7\pi^5f_\pi^4}\approx 2,6\times 10^{-2}pb,\nn
I=\frac{1}{4}\int\limits_0^\infty\frac{d z}{z^4}\ln^8(\sqrt{1+z}+\sqrt{z})= 0.3557.
\ea
Thus the   expressions (9),  (12),  (13)  allow  to calculate the yields of any combination of light meson pairs produced  in $\gamma\gamma$ collisions.
\section{Bound states production}
The considered   approach is especially efficient   in investigation  of  bound states formation  in  collisions of high energy particles.
As a typical examples we examine the production of  simplest  atoms being the bound state of two charged  pions  (pionium atom $A_\pi$) and atom constructed from two fermions  (positronium atom  Ps). \\
To determine the pionium impact  factor   $\gamma\gamma^\ast\to A_\pi$  we take advantage of  well known QED  amplitude~\cite{AB59}  for  the process $\gamma(k_1,e_1)+\gamma^\ast(q) \to \pi^-(q_-)+\pi^+(q_+)$:
\ba
M^{\gamma\gamma ^\ast\to \pi\pi}=\frac{4\pi\alpha}{s} [\frac{(2q_-e_1) ((-2q_+ +q)k_2)}{2q_-k_1}+
\frac{(-2q_+e_1) ((2q_- -q)k_2)}{-2q_+ k_1}-2(e_1k_2)]\nn
\ea
Account on  that in atom   pions have almost the  same velocity $q_+=q_-=p/2$;  $2(p k_1)=4m_\pi^2+\vec q^2$ and  expressing $p, e$ through the Sudakov variables:
\ba
p=\alpha_p k_2+\beta_p k_1+q_\bot;  ~~~e=\beta_e k_1+e_\bot,
\ea
Making use the relation:
\ba
(pe_1)((p-q)k_2-(pk_1)(e k_2)=-2s(\vec{q}\vec{e}_1)
\ea
 the amplitude of two pions production with the same velocities takes the form:
\ba
M^{\gamma\gamma^\ast \to \pi\pi}=\frac{8\pi\alpha(\vec e\vec q)}{\vec q^2+4m_\pi^2}
\ea
In order to obtain the amplitude for pionium production  we use the   relation~\cite{staffin77} allowing to connect the amplitude of  two free scalar mesons production with the  production amplitude of  their bound state $A_\pi$ \footnote{The square of pionium ground state wave function at origin has the form: $|\Psi(0)|^2=\frac{\alpha^3 m^3}{8\pi}$}
\ba
M^{\gamma\gamma^\ast \to A_\pi}=M^{\gamma\gamma^\ast \to \pi\pi}\frac{i\Psi(\vec r=0)}{\sqrt{m_\pi}}
\ea
Finally  for  the amplitude of  pionium production in  $\gamma\gamma^\ast$ collisions we get:
\ba
M^{\gamma\gamma^\ast \to A_\pi}=\frac{8i\pi\alpha (\vec e\vec q)}{4m_{\pi}^2+\vec{q}^2}\frac{\Psi(0)}{\sqrt{m_\pi}},
\ea
To obtain the  impact factor  for  para-positronium creation  $ \gamma(k_1,e_1)+\gamma^\ast (q) \to Ps(p)$  we take advantage  of the  receipt ~\cite{novikov78, gevorkyan98} of passage from free $e^+e^-$ pair to their bound state   and textbook  expression~\cite{AB59} for $e^+e^-$
pair creation in $\gamma\gamma$ collisions. As a result the matrix element for bound state creation takes the form:
\ba
M^{\gamma\gamma\ast\to Ps}&=&i\frac{4\pi\alpha}{s}\frac{m_e\sqrt{\alpha^3}}{\sqrt{4\pi }}
\frac{1}{4}Tr[\hat{e}_1\frac{\hat{q}_--\hat{k}_1+m_e}{(q_- -k_1)^2-m_e^2}\hat{k}_2\nn
&+&\hat{k}_2\frac{-\hat{q}_+ +\hat{k}_1+m_e}{(-q_+ +k_1)^2-m_e^2}\hat{e}_2](\hat{p}+m_{Ps})\gamma_5.
\ea
Making use the relations  $q+ k_1=p=q_+ +q_-$ and $q_+=q_-=p/2$ one   obtains:
\ba
M^{\gamma\gamma^\ast \to Ps}=\frac{4im_e\sqrt{\pi\alpha^5}}{4m_e^2+\vec{q}^2}|[\vec{e}_1,\vec{q}]|.
\ea
With the help of equation  (9) and impact factors  (23),  (25)  one  can  calculates  the differential cross section  of elementary atoms creation in the processes:
\ba
\gamma+\gamma\to S_1+S_2;~~~~S_1, S_2=A_\pi,  Ps.
\ea
For reader convenience and rough estimates  of the order  of total cross sections of  bound state production by photon exchange mechanism (Fig.1a)
 we cite the expressions for the total cross sections relevant to reactions (26):
\ba
\sigma^{\gamma\gamma\to Ps Ps}=\frac{\pi\alpha^8}{96}r_e^2(1+2\cos^2\phi_0);~~~~  \sigma^{\gamma\gamma\to A_\pi A_\pi}= (\frac{r_e}{4r_\pi})^2\sigma^{\gamma\gamma\to Ps Ps}; \nn
\sigma^{\gamma\gamma\to Ps A_\pi}=\frac{\pi\alpha^8}{64}r_\pi^2(3-2\cos^2\phi_0);~~~
\sigma^{\gamma\gamma\to \pi^0 Ps }=\frac{\alpha^7}{32\pi^2f_\pi^2}(1+2\cos^2\phi_0);\nn
r_e=\frac{\alpha}{m_e}, r_\pi=\frac{\alpha}{m_\pi}.
\ea
Rough estimates of these cross sections  show  that they are really very small quantity  of order  $10^{-8}nb$.

\section{Vector meson exchange in  pairs production}
Up to now we considered production processes provided by photon exchanges  (Fig.1a).  From the other hand the vector meson (Fig. 1b) exchanges also give  nondecreasing with energy contribution to  the processes  (1).  The problem with such type exchanges is connected with the fact that the Born approximation depicted on  Fig. 1b  badly violated for strong interactions. \\
To take  into account the higher order contributions of strong interaction one would   replaces   the exchanged vector meson propagator  in (9)  by  its reggeized  analog ~\cite{laget97}
\ba
\frac{1}{t-m_V^2}\to\alpha'\frac{1-e^{-i\pi\alpha(t)}}{2}\Gamma {(1-\alpha(t))}(\frac{s}{s_0})^{\alpha(t)},
\ea
where $\alpha(t)$  is the Regge trajectory of vector meson
\ba
\alpha(t)=\alpha(0)+\alpha' t
\ea
 The $\Gamma$ function contains the pole propagator $sin^{-1}(\pi\alpha(t))$  and in the limit $t\to m_V^2$ the expression (28) reduced to the standard pole propagator.  The detailed characteristics  of  Regge trajectories  of different vector mesons can be found in work \cite{titov08} and references therein.\\
Later on for  estimation of  vector mesons  contribution to the relevant cross sections we use  the simplified suppression factor:
\ba
R(s,t)=(\frac{s}{s_0})^{2(\alpha(t)-1)} \approx \frac{s_0}{s};~~ s_0\approx 1GeV^2.
\ea
The impact factors  corresponding to the vector mesons exchanges depend on the considered process and would  be obtained   as it has been done above for  photon exchanges. \\
As an example  let us consider the  process of two charged pions  production  $\gamma\gamma\to \pi^+\pi^-$   for which the photon exchange is absent.  The main contribution to this reaction at high energies  gives the  $\rho$  exchange.\footnote{The pion exchange~\cite{schmitz94} dominates only at small transfer momenta $t\leq 4m_\pi^2\leq 0.1GeV^2$ and fall off with energy much stronger than vector exchanges.}  The matrix element of radiative decay of charged meson  $\rho^+(p,e_1)\to \pi^+(p_\pi)+\gamma(k,e_2)$  reads  $M=g_+(pe_1 k e_2)$,  where the  constant $g_+$ can be determined from the relevant decay  width :
\ba
\Gamma^{\rho^+\to\pi^+\gamma}=\frac{g_+^2}{96\pi}(\frac{m_\rho^2-m_\pi^2}{m_\rho})^3.
\ea
Comparing this relation  with the experimental value of the  $\rho^+\to\pi^+\gamma$   branching ratio~\cite{pdg}  $B=4.5\times 10^{-4}$, $\Gamma=67keV$   one gets  $g_+\approx 0.21 GeV^{-1}$.\\
In  the case when one of the photons is  virtual  it is enough to do the  simple replacement  $g_+\to g_+F(z)$ with
\ba
F(z)=\frac{4}{z}\ln^2(\sqrt{1+\frac{z}{4}}+\sqrt{\frac{z}{4}});~~~ z=\frac{\vec{q}^2}{m_q^2}.
\ea
The differential cross section  of the process $\gamma\gamma\ast\to\pi^+\pi^-$ in peripheral kinematic takes the form
\ba
d\sigma &=&\frac{d\vec{q}^2 d\phi}{32\pi^2}\frac{|M^{(1)}|^2|M^{(2)}|^2}{(\vec{q}^2+m_\rho^2)^2}, \nn
M^{(1)}&=&\frac{g_+}{2}[\vec{q}\vec{e}_1]F(z);~~~M^{(2)}=\frac{g_+}{2}[\vec{q}\vec{e}_2]F(z).
\ea
Averaging over azimuthal angle according to equation (14)  for the total cross section we obtain:
\ba
\sigma (\gamma\gamma\to \pi^+\pi^-)=\frac{g_+^4m_q^2}{32\pi}(1+2\cos^2\phi_0)I;\nn
I=\int\limits_0^\infty\frac{d z}{z^2(z+(\frac{m_\rho}{2m_q})^2)^2}\ln^8(\sqrt{1+\frac{z}{4}}+\sqrt{\frac{z}{4}})\approx 0.372 \nn
\sigma^{\gamma\gamma\to\pi^+\pi^-}\approx 60 (1+2\cos^2\phi_0)(\frac{s_0}{s}) nb.
\ea
In the same way one can estimates the contribution from   $\rho,\omega$ exchanges to the process of  two neutral pions production
$\gamma\gamma\to \pi^0\pi^0$, determining the  constants   $g_\rho, g_\omega$  from  experimental data on the relevant decay rates \cite{pdg}
\ba
\Gamma(\rho^0 \to \pi^0\gamma)&=&8.9\times10^{-5}GeV;~~~g_\rho=0.25GeV^{-1}\nn
\Gamma(\omega \to \pi^0\gamma)&=&70\times10^{-5}GeV;~~~g_\omega=0.71GeV^{-1}.
\ea
For the total  cross section  of the  process  $\gamma\gamma\to\pi^0\pi^0$  provided by vector meson exchanges we obtain:
\ba
\sigma^{\gamma\gamma\to \pi^0\pi^0}=3 (1+2\cos^2\phi_0)(\frac{s_0}{s})\mu b.
\ea
\section{Pionium atom production in $ep$ and $\pi p$ collisions.}
In recent years there has been a significant effort to extract the $\pi\pi$ s-wave scattering lengths $a_I$ with total isospin I=0, 2 from experimental data on pionium atom  $A_\pi$ creation. The  scattering lengths determination with high precision allows to check the predictions of low-energy hadron theories such as Chiral Perturbation Theory (CHPT) or Nambu-Jona-Lasinio model (NJL) which give it value with unprecedented for strong interaction accuracy $\sim 2\%$~\cite{colangelo01}.\\
The main goal of experiment DIRAC~\cite{dirac11} at PS CERN has been the determination of pions scattering lengths difference $a_0-a_2$ from the measurement of pionium atom lifetime, which is connected with this difference  by the relation~\cite{uretsky61}:
\ba
\Gamma=\frac{1}{\tau}=\frac{2}{9}\sqrt{\frac{2(m_{\pi^+}-m_{\pi^0})}{m_\pi}}(a_0^0-a_0^2)^2m_\pi^3\alpha^3.
\ea
At present due to experiment Dirac and experiments on  kaons decays~\cite{batley09,gevorkyan07} the scattering lengths determined from experimental data with precision comparable with theoretical predictions.\\
Below we will consider the peripheral mechanism of creation of two charged pions in collision
of high energy electron with the proton and similar one with the initial high energy negatively charged
$\pi$-meson instead electron
\ba
e(p_1)+p(p_2) \to e(p_1')+A_{\pi}(p)+p(p_2')
\ea
\ba
\pi(p_1)+p(p_2) \to \pi(p_1')+A_{\pi}(p)+p(p_2')
\ea
$s=(p_1+p_2)^2>> m_p^2$  with $m_p$  a  proton mass.\\
For the case of electron-proton collision the pion pair is created in the collision of virtual photon emitted by electron and virtual $\rho$ ($\omega$) meson emitted by proton (Fig. 2a). In the case of $\pi$-meson proton interaction  the pion pair  is produced by two virtual $\rho$ mesons (Fig. 2b).
\begin{figure}[ht]
\begin{center}
\includegraphics[scale=0.5]{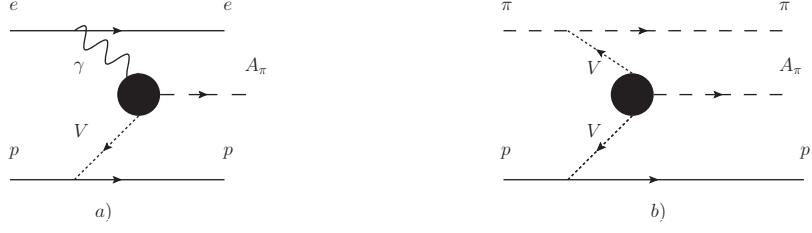}
\caption{ a) The pionium  electroproduction  in the process $ e+p\to e+p+A_\pi $ ;  b) Pionium production by pions $ \pi+p\to \pi+p+A_\pi$}
\end{center}
\end{figure}
The matrix element corresponding to  these processes  has the form:
\ba
M=\frac{G}{(q_1^2-m_1^2)(q_2^2-m_2^2)}J_1(p_1)_{\mu_1}T_{\mu\nu}J_p(p_2)_{\nu_1} G^{\mu\mu_1}G^{\nu\nu_1},
\ea
with $G$ is the product of the relevant coupling constants, $m_{1,2}$-masses of the exchanged vector
particles;  $J_1, J_p$  are the currents  connected with the colliding  particles;  tensor $T_{\mu\nu}$ describes the
conversion of two vector mesons to pion pair.\\
The main contribution in peripheral kinematics (non-vanishing in limit $s\to \infty$) arises from relevant Green functions:
\ba
G^{\mu\mu_1}=\frac{2}{s}p_2^\mu p_1^{\mu_1};G^{\nu\nu_1}=\frac{2}{s}p_2^\nu p_1^{\nu_1}.
\ea

Matrix element of the sub-process of creation of pion pair with equal 4-momenta by two virtual vector
particles
\ba
V_\mu(q_1)+V_\nu(q_2) \to \pi^+(q)+\pi^-(q)
\ea
is described by the  tensor:
\ba
T_{\mu\nu}=\frac{2}{D}[q_{2\mu}q_{1\nu}+Dg_{\mu\nu}], D=-\frac{1}{2}[4m^2+\vec{q}_1^2+\vec{q}_2^2],
\ea
Combining these expressions one gets   for the  matrix element of the process $e+p\to e'+p'+A_\pi$:
\ba
M^{e p\to e p A_\pi}=\frac{4s}{\vec{q}_1^2+m_e^2\beta_1^2}\frac{G_e}{\vec{q}_2^2+m_V^2}\Phi_e\Phi_A\Phi_p\Psi(0),
\ea
where  $G_e=4\pi\alpha g_\pi g_p$, ($g_\pi,g_p$-coupling constants of $\rho$-meson with pion and proton, which we put  $g_\pi=g_\rho=3$)
\ba
\Phi_e=\frac{1}{s}\bar{u}(p_1')\hat{p}_2u(p_1); ~~\Phi_A=\frac{1}{s}p_1^\mu p_2^\nu T_{\mu\nu}=-2\frac{\vec{q}_1\vec{q}_2}{D}; \nn
\Phi_p=\frac{1}{s}\bar{u}(p_2')\Gamma_\mu u(p_1)p_2^\mu, \Gamma_\mu=\gamma_\mu F_1+\frac{1}{4M_p}(\hat{q}_2\gamma_\mu-\gamma_\mu\hat{q}_2)F_2,
\ea
Here $F_1=F_1(q_2^2), F_2=F_2(q_2^2)$ are Dirac and Pauli form-factors of proton.\\
The phase volume of the three particles in the  final state:
\ba
d\Gamma=\frac{(2\pi)^4}{(2\pi)^9}\frac{d^3p_1'}{2E_1'}\frac{d^3p_2'}{2E_2'}\frac{d^3p_A}{2E_A}\delta^4(p_1+p_2-p_1'-p_2'-p_A),
\ea
can be reduced using the Sudakov variables to the following form:
\ba
d\Gamma=\frac{1}{(2\pi)^5}\frac{1}{4s}\frac{d\beta_1}{\beta_1}d^2\vec{q}_1d\vec{q}_2.
\ea
Making use the summed over  spin states of the squares of matrix elements of the relevant  sub-processes:
\ba
\sum|\Phi_e|^2=2;~~\sum|\Phi_p|^2=2[F_1^2+\frac{\vec{q}_2^2}{4m_p^2}F_2^2]; \nn
|\Phi_A|^2=\frac{4(\vec{q}_1\vec{q}_2)^2}{(4m^2+\vec{q}_1^2+\vec{q}_2^2)^2},
\ea
where m is the pion mass. The cross section of the process  $e+p \to e+p+A_\pi$ takes the form:
\ba
d\sigma^{e p \to e p A_\pi}&=&\frac{\alpha^5g_\pi^2g_p^2}{2\pi^2}\frac{m^2 \vec{q}_1^2 d\vec{q}_1^2\vec{q}_2^2 d\vec{q}_2^2}{(4m^2+\vec{q}_1^2+\vec{q}_2^2)^2(\vec{q}_1^2+m_e^2\beta_1^2)^2
(\vec{q}_2^2+m_\rho^2)^2}\nn &\times&[F_1^2+\frac{\vec{q}_2^2}{4m_p^2}F_2^2]\frac{d\beta_1 (1-\beta_1)}{\beta_1};~~~
\frac{4m^2}{s}<\beta_1<1.
\ea
Similar expression for the cross section with initial $\pi$ meson instead of the electron:
\ba
d\sigma^{\pi p \to \pi p A_\pi}&=&\frac{\alpha^3g_\pi^6g_p^2}{64\pi^4}\frac{m^2 \vec{q}_1^2 d\vec{q}_1^2\vec{q}_2^2 d\vec{q}_2^2}{(4m^2+\vec{q}_1^2+\vec{q}_2^2)^2(\vec{q}_1^2+m_\rho^2)^2
(\vec{q}_2^2+m_\rho^2)^2}\nn &\times&[F_1^2+\frac{\vec{q}_2^2}{4m_p^2}F_2^2]\frac{d\beta_1 (1-\beta_1)}{\beta_1}.
\ea
Integrating these expressions over phase volume one obtains the total yield of pionium atom. In the case of the electroproduction:
\ba
\sigma(ep\to epA_\pi)=\sigma_e D_e,~~ \sigma_e=\frac{\alpha^5g_\pi^2g_p^2m^2}{2\pi^2m_\rho^4}\approx 0.3 pb ;\nn
D_e=J_N[l_m^2+l_\pi(l_m-1)-2], J_N=\int\limits_0^\infty \frac{x N^2 d x}{(x+4)^2(x+N)^2}\approx 0.845; \nn
l_m=\ln\frac{s}{4m^2},~~~l_\pi=\ln\frac{m^2}{m_e^2}.
\ea
For $s=100 GeV^2$  the cross section $\sigma(ep\to ep A_\pi) \approx 30 pb$ is too small to be measured at present accelerators. \\
As to the pionium production by pions we obtain:
\ba
\sigma(\pi p\to \pi p A_\pi)=\sigma_\pi D_\pi,~~~ \sigma_\pi=\frac{\alpha^3g^8m^2}{64\pi^2m_\rho^4}\approx 217nb; \nn
D_\pi=(l_m-1)I, I=\int\limits_0^\infty\int\limits_0^\infty\frac{x_1x_2 d x_1 d x_2}{(x_1+x_2)^2(x_1+1)^2(x_2+1)^2}\approx 0.133.
\ea
 The total cross section turns out to be of the order $\sigma(\pi p\to \pi p A_\pi)\approx 178 nb$ for  s=80 $GeV^2$ (IHEP, Protvino) and thus  can be measured at modern facilities.\\
In conclusion we note that the contribution from  the channels with exchange of two photons is of order
\ba
\sigma_e^{\gamma\gamma}=\sigma_0 D_e^{\gamma\gamma};~~\sigma_\pi^{\gamma\gamma} =\sigma_0 D_\pi^{\gamma\gamma}, \sigma_0=\frac{8\alpha^7}{m^2}\approx 1,8 \times 10^{-3}pb.
\ea
In spite of a rather large enhancement factors $D_e^{\gamma\gamma}\sim 10 D_\pi^{\gamma\gamma}\sim 10^2$  the relevant contributions can  be safely neglected.
\section{The vector meson exchange reggeization}
As was mentioned above the consideration of hadronic processes  in peripheral kinematics in Born approximation is non-adequate. The effect of converting the ordinary vector mesons to the relevant Regge poles must be taken into account. It results in an additional suppression  factor to the total cross sections of processes (38), (39):
\ba
R=\left(\frac{s_1s_2}{s_0^2}\right)^{2(\alpha(0)-1)},
\ea
Keeping in mind the kinematical relation $s_1s_2\approx 4m^2 s$ and puting $\alpha(0)\approx 0.5$ :
\ba
R\approx \frac{s_0^2}{4sm^2}.
\ea
For instance at $s=80GeV^2$ it results in the suppression  factor
\ba
R\approx 0.16.
\ea
So the realistic cross section for this energies  is about $\sigma^\pi\approx 28 nb$.\\
Let us note that in the double pomeron exchanges  in the  process (39)  (or pionium photoproduction off pomeron in the case of reaction (38)) such suppression factor is absent and at enough high energies the pomeron exchanges dominated.
It is useful to estimate the energies from which the photon exchange becomes comparable with vector mesons one.
For instance to obtain the matrix element  for pionium  electroproduction   by two photon exchanges from the matrix element with one vector meson exchange  (fig.2a) it is enough  to do a simple replacement:
\ba
g_\pi g_p\frac{s_0}{2m\sqrt{s}}\to 4\pi\alpha
\ea
Thus only from  energies $s\sim 10^5 GeV^2$  the contribution with  two photon exchanges  in pionium electroproduction  becomes larger than the one with  vector meson exchange.

\section{Acknowledgements}
Authors are grateful to A. Ahmadov, N. Kochelev and R. Togoo for discussions.
The work of E. Kuraev was partially supported by RFBR-01201164165  and  Belorussian grants.


\Bibliography{99}
\bibitem{telnov06} V. I. Telnov, Acta Physica Polonica {\bf B37}, 1049 (2006)
\bibitem{arutjunian63} F. R. Arutjunian , V. A. Tumanyan,  ZETP, {\bf 44}, 2100 (1963)
\bibitem{serbo10} V. G.  Serbo, Nucl. Instr. Meth. , {\bf 472}, 260 (2010)
\bibitem{lep1} ALEPH Collaboration, Phys. Lett.{\bf B569} 140 (2003)
\bibitem{lep2} L3 Collaboration, Phys. Lett.{\bf B615} 19 (2005)
\bibitem{kek1} Belle Collaboration, hep-ex/0711.1926
\bibitem{weinberg79} S. Weinberg, Physica, {\bf A96}, 327 (1979)
\bibitem{gasser85} J. Gasser, H. Leutwyler, Nucl. Phys., {\bf B250}, 465 (1985)
\bibitem{volkov05} M. K. Volkov, A. E. Radzhabov, arXiv: hep-ph/0508263
 \bibitem{arbuzov10}  A. B. Arbuzov  et al.,  Particles and Nuclei, {\bf 41},  1113  (2010).
\bibitem{AB59} A. I. Akhiezer, V. B.  Berestetskij,  Quantum Electrodynamics, Moscow, 1959.
\bibitem{staffin77} R. Staffin, Phys. Rev.{\bf D16}, 726 (1977)
\bibitem{novikov78} V. A. Novikov   et al., Phys. Rep. {\bf 41C} (1978)
\bibitem{gevorkyan98} S. R.  Gevorkyan  et al. , Phys. Rev. {\bf A 58}, 4556 (1998)
\bibitem{laget97} M. Guidal, J. M. Laget, M. Vanderhaeghen, Nucl. Phys. {\bf A627} 645 (1997)
\bibitem{titov08} A. V. Titov,  B.  Kampfer  arXiv: hep-ph/0807.1822
\bibitem{schmitz94} N. Schmitz, Nucl. Phys. {\bf B 36},145 (1994)
\bibitem{pdg}C. Amsler et al. (PDG),  Phys. Lett. {\bf B667}, 1 (2008)
\bibitem{colangelo01} G. Colangelo, J. Gasser, H. Leutwyler, Nucl. Phys. {\bf B603}, 125 (2001)
\bibitem{dirac11}B. Adeva et al., Phys. Lett.  {\bf B704}, 24 (2011)
\bibitem{uretsky61} J. Uretsky, J.Palfrey, Phys. Rev. {\bf 121}, 1798 (1961)
\bibitem{batley09} J. R. Batley et al.,  Eur. Phys. J. {\bf C64}, 589  (2009)
\bibitem{gevorkyan07}S. R. Gevorkyan, A.V. Tarasov,  O. O. Voskresenskaya, Phys.Lett. {\bf  B649}, 159 (2007)
\endbib
\end{document}